# Cooling with Superfluid Helium

*Ph. Lebrun[1] and L. Tavian*
CERN, Geneva, Switzerland

**Abstract**
The technical properties of helium II ('superfluid' helium) are presented in view of its applications to the cooling of superconducting devices, particularly in particle accelerators. Cooling schemes are discussed in terms of heat transfer performance and limitations. Large-capacity refrigeration techniques below 2 K are reviewed, with regard to thermodynamic cycles as well as process machinery. Examples drawn from existing or planned projects illustrate the presentation.

## 1    Introduction

Once confined to low-temperature physics laboratories, superfluid helium[2] has become a technical coolant for advanced superconducting devices, to the point that it is now implemented in industrial-size cryogenic systems, routinely operated with high reliability. There are two classes of reason that call for the use of superfluid helium as a coolant for superconducting devices; namely, the lower temperature of operation, and the enhanced heat transfer properties at the solid/liquid interface and in the bulk liquid.

The lower temperature of operation is exploited in high-field magnets [1, 2], to compensate for the monotonously decreasing shape of the superconducting transition frontier (the 'critical line') in the current density versus magnetic field plane, shown in Fig. 1 for some superconducting materials of technical interest. In this fashion, the current-carrying capacity of the industrial Nb–Ti superconducting alloys can be boosted at fields in excess of 8 T, thus opening the way for their use in high-field magnet systems for condensed-matter physics [3–5], nuclear magnetic resonance [6, 7], magnetic confinement fusion [8, 9], and circular particle accelerators and colliders [10–12]. In the case of high-frequency superconducting devices such as acceleration cavities [13], the main drive for superfluid helium cooling is the exponential dependence of the BCS losses on the ratio of the operating temperature to the critical temperature. Accelerators based on this technology, such as medium-energy, high-intensity machines [14–16], electron linacs feeding free-electron lasers [17], and future high-energy lepton colliders [18–20], operate in a temperature range that minimizes the capital costs and the overall energy consumption. This issue is schematized in Fig. 2.

The technical heat transfer characteristics of superfluid helium basically derive from peculiar transport properties [21, 22]. Its low bulk viscosity enables superfluid helium to permeate to the heart of magnet windings, while its very large specific heat (typically $10^5$ times that of the conductor per unit mass, $2 \times 10^3$ per unit volume), combined with excellent heat conductivity at moderate heat flux ($10^3$ times that of cryogenic-grade OFHC copper) can produce powerful stabilization against thermal disturbances. In order to fully exploit these properties in both steady-state and transient regimes – for example, for power heat transport over macroscopic distances as well as intimate stabilization of superconductors inside magnet windings – an elaborate thermo-hydraulic design of the cooling circuits, conductor, insulation, and coil assemblies is required. This often conflicts with other technical or economic requirements of the projects and acceptable trade-offs have to be found.

---

[1] philippe.lebrun@cern.ch
[2] Strictly speaking, we are referring to the second liquid phase of helium, called He II, which exhibits the unusual bulk properties associated with superfluidity and is therefore also called a 'superfluid'. This is not to be confused with the entropy-less component of the phenomenological two-fluid model accounting for the behaviour of He II, for which some authors prefer to retain the term 'superfluid'.

In the following, we will only address the specific issues of cryogenic technology pertaining to the use of superfluid helium as a technical coolant, namely different cooling methods as well as processes and machinery for sub-lambda temperature refrigeration [23]. Reference is made to companion lectures for cryogenic techniques which – however important in system design – are not superfluid-helium specific, such as conventional heat transfer and cryostat design [24, 25].

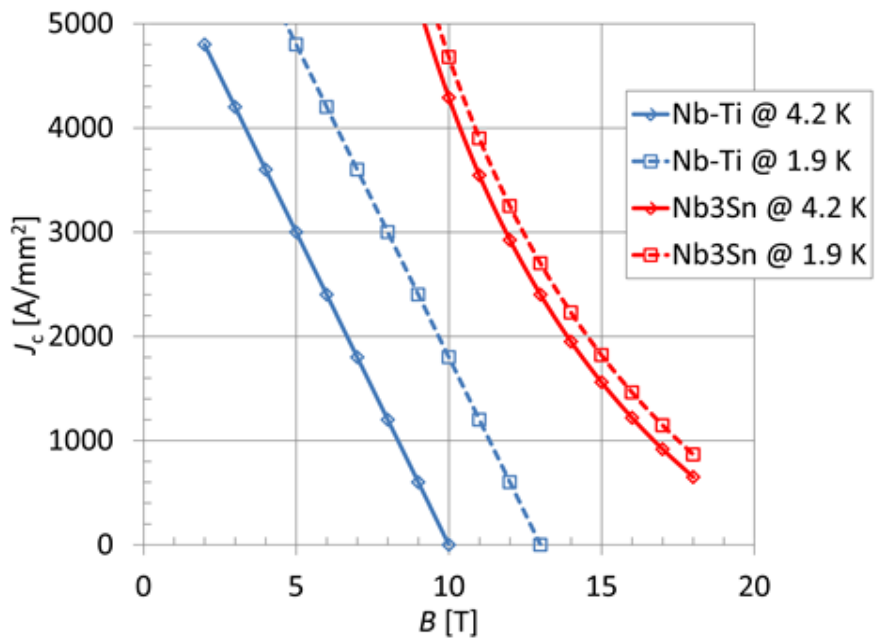

**Fig. 1:** The critical current density of technical superconductors

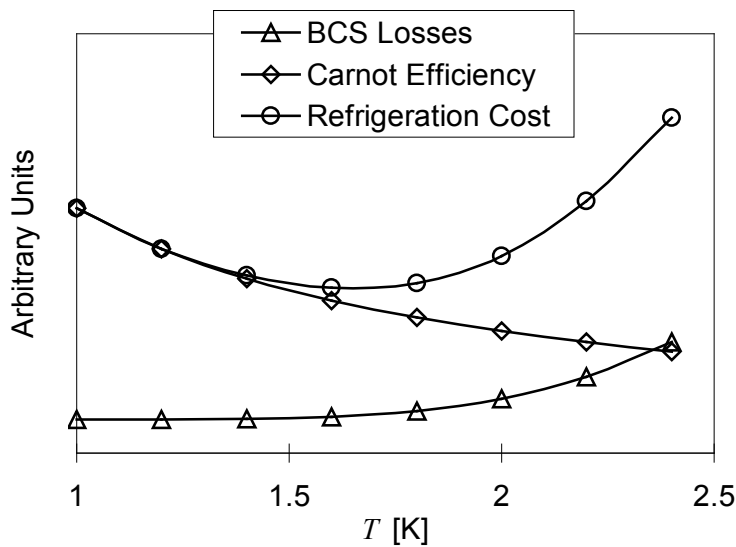

**Fig. 2:** The optimal operating temperature of RF superconducting cavities

## 2 Different cooling methods

### 2.1 Pressurized versus saturated superfluid helium

A glance at the phase diagram of helium (Fig. 3) clearly shows the working domains of saturated helium II, reached by gradually lowering the pressure down to below 5 kPa along the saturation line, and pressurized helium II, obtained by sub-cooling liquid helium at any pressure above saturation, and in particular at atmospheric pressure (about 100 kPa).

Although requiring one more level of heat transfer and additional process equipment – in particular, a pressurized-to-saturated helium II heat exchanger – pressurized helium II cooling confers several important technical advantages [26]. The avoidance of low-pressure operation in large and complex cryogenic systems clearly limits the risk of air in-leaks, and resulting contamination of the process helium. Moreover, in the case of electrical devices, the low dielectric strength exhibited by low-pressure helium vapour [27] in the vicinity of the minimum of the Paschen curve (Fig. 4) [28] leads to the additional risk of electrical breakdown at fairly low voltage. Operating in pressurized helium II avoids this kind of problem.

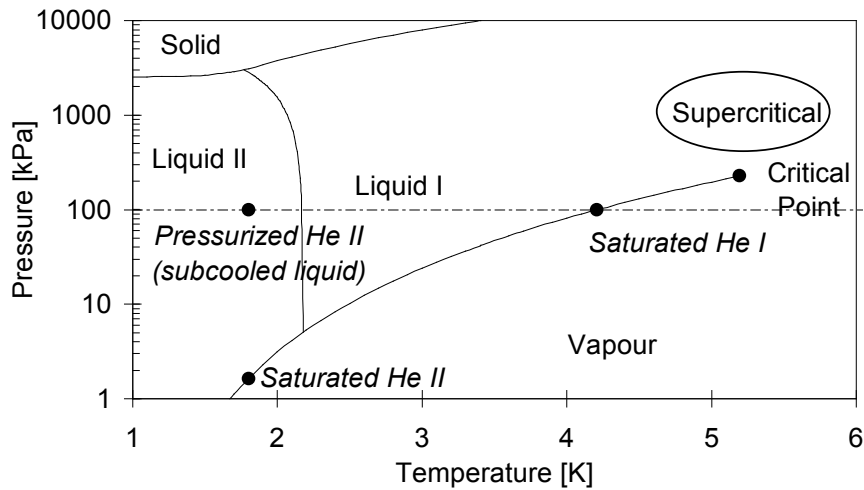

**Fig. 3:** The phase diagram of helium

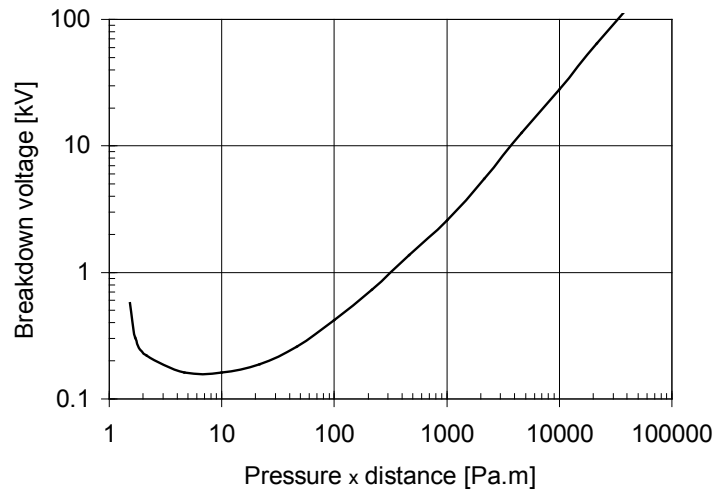

**Fig. 4:** The Paschen curve for helium at 300 K

However, the most interesting and specific aspect of pressurized helium II in the operation of superconducting devices stems from its capacity for cryogenic stabilization. As a sub-cooled (monophase) liquid with high thermal conductivity, pressurized helium II can absorb in its bulk a deposition of heat, up to the temperature at which the lambda line is crossed, and local boiling starts only then, due to the low thermal conductivity of helium I. Quasi-saturated helium II, which is in fact slightly sub-cooled due to the hydrostatic head below the surface of the liquid bath, may only absorb heat deposition up to the point at which the saturation line is crossed and change of phase occurs. The enthalpy difference from the working point to the transition line is usually much smaller in the latter case. The argument, developed in Ref. [29], typically yields an order of magnitude better performance in favour of pressurized helium II.

## 2.2 Bath cooling

The first attempts to lower the temperature of a helium bath at atmospheric pressure were made with the purpose of operating superconducting magnets at temperatures below 4.2 K [30]. Thermal stratification in the 'Roubeau' bath (Fig. 5) permitted the lambda point to be reached, but the large heat exchange area with the sub-cooled helium I volume, combined with the high thermal conductivity of the helium II bath, prevented the latter from cooling down further. By inserting a restriction to thermal conduction, in the form of an insulating plate with a minimum residual helium cross-section, the 'Claudet' bath [31] enabled the helium II volume to be sub-cooled well below the lambda point, to a temperature at which its heat transport properties are maximal. Both systems need an external source of refrigeration below 2 K, usually in the form of a saturated helium II heat exchanger coupled to the pressurized bath.

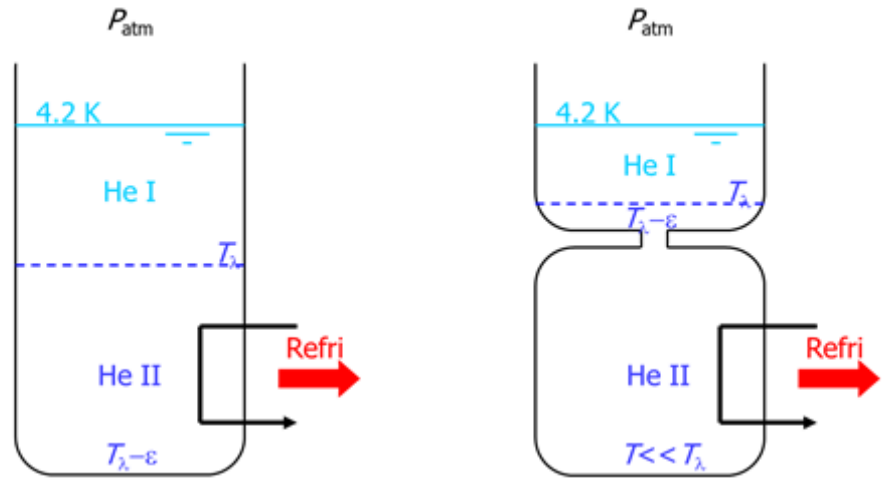

**Fig. 5:** Bath cooling with pressurized helium II: the Roubeau bath (left) and the Claudet bath (right)

## 2.3 Conduction cooling

In the following, we shall only consider conductive heat transport in helium II at heat fluxes of technical interest (typically above 1 kW·m$^{-2}$). For most practical geometries, this means working in the 'turbulent' regime, with full mutual friction between the components of the two-fluid model [32]. In this regime, helium II exhibits a large, finite, and non-linear bulk heat conductivity, the value of which depends both on temperature and heat flux. While the general patterns of this behaviour can be predicted by the Gorter–Mellink [33] theory,[3] practical data useful for engineering design has been established in a number of experiments [34–39].

---

[3] In 1949, C.J. Gorter and J.H. Mellink introduced the idea of an interaction producing mutual friction between the components of the two-fluid model, to account for the observed transport properties of helium II.

Consider conduction in one dimension; for example, in a tubular conduit of length $L$, the ends of which are maintained at temperatures $T_C$ and $T_W$. The steady-state heat flux $\dot{q}$ is given by

$$\dot{q}^n L = X(T_C) - X(T_W), \qquad (1)$$

where the best experimental fit for $n$ is 3.4, and $X(T)$ is a tabulated function of temperature, physically analogous to a conductivity integral [34]. A plot of this function reveals that the apparent thermal conductivity of helium II goes through a maximum at around 1.9 K (Fig. 6).

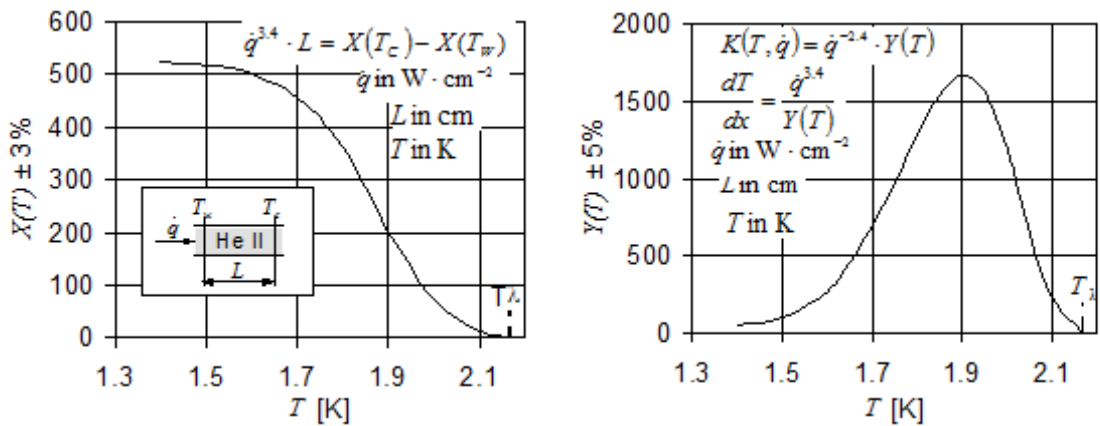

**Fig. 6:** The thermal conductivity integral and apparent thermal conductivity of pressurized superfluid helium [34].

As an example, the heat flux transported by conduction between 1.9 K and 1.8 K in a 1-m long static column of helium II is about 1.2 W·cm$^{-2}$; that is, three orders of magnitude higher than would be conducted along a bar of OFHC copper of the same geometry! The non-linearity with respect to heat flux also results in a much weaker dependence of conduction upon length or thermal gradient. Figure 7 shows the steady-state conduction $\dot{Q}$ in superfluid helium between 1.9 K and 1.8 K versus the static column length $L$ for different equivalent nominal diameters of the column. This abacus clearly shows that while the heat flux conducted in a solid is directly proportional to the thermal gradient applied, doubling the conduction length in a column of helium II only reduces the heat flux by some 20%.

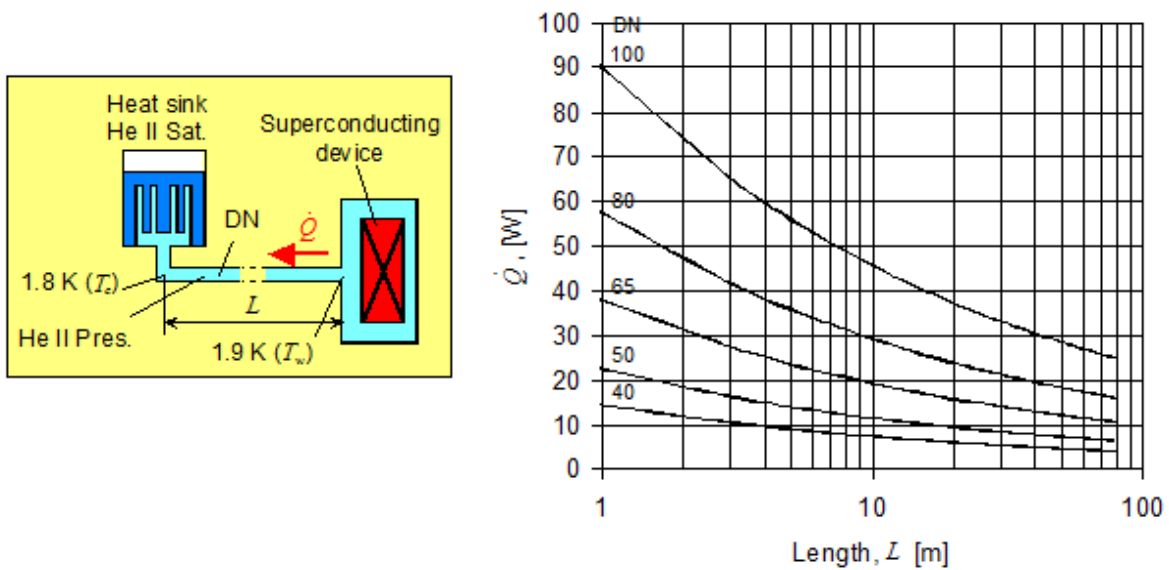

**Fig. 7:** Steady-state conduction in pressurized superfluid helium

The variation of *X(T)* also implies that, for each value of the cold boundary temperature $T_C$, there exists a maximum possible heat flux at which $T_W$ reaches the lambda point, and the helium column ceases to be superfluid. Values of this limiting heat flux, which also weakly depends on *L*, range from a fraction to a few units of W·cm$^{-2}$, for practical cases of interest. This clearly places an intrinsic limitation on the applicability of helium II conduction for quasi-isothermal cooling of long strings of superconducting devices in an accelerator. Transporting tens of watts over distances of tens of metres would then require a temperature difference of several hundred millikelvins and a large cross-section of helium, which is both impractical and thermodynamically costly. For a more precise estimate, consider a uniformly heated tubular conduit of length *L*, operating between temperatures $T_C$ and $T_W$, and apply the helium II steady-state conduction equation to this fin-type geometry. After integration,

$$\dot{q}_{total}{}^n L = (n+1)\left[X(T_C) - X(T_W)\right], \quad (2)$$

where $\dot{q}_{total}$ is the total heat flux flowing through the section at temperature $T_C$, near the heat sink. Figure 8 shows the steady-state conduction $\dot{Q}_{tot}$ in superfluid helium of a cryomagnet string with linear heating $\xi$ between 1.9 K (the temperature of the warmest magnet) and 1.8 K (the temperature at the heat sink). As an example, the cooling by conduction of a 50-m long cryomagnet string, with a uniform linear thermal load of 1 W·m$^{-1}$, would require a helium II cross-section of 90 cm$^2$; that is, a 10.7-cm diameter conduit. In view of such constraints, the conduction-cooling scheme originally considered for the LHC project [40] was later abandoned, in favour of the more efficient one described in Section 2.5.

Conduction through static pressurized superfluid helium, however, remains the basic process for extraction and local transport of heat from the LHC magnet windings, across their polyimide-wrap electrical insulation. Although the polyimide tape, which constitutes the insulation of the superconducting cable, is wrapped in two layers with a half overlap (Fig. 9) in order to achieve sufficient mechanical toughness and dielectric strength, this still preserves sufficient percolation paths for helium II conduction to significantly improve the heat transfer, well above the solid conduction across the sole polyimide [41].

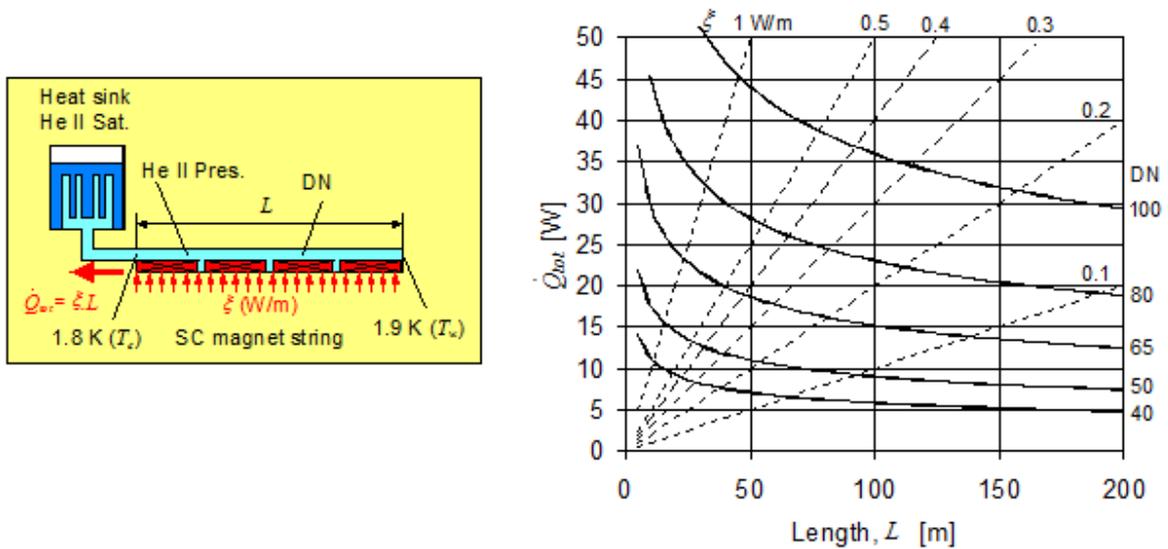

**Fig. 8:** The steady-state conduction cooling of a cryomagnet string with a linear applied heat load

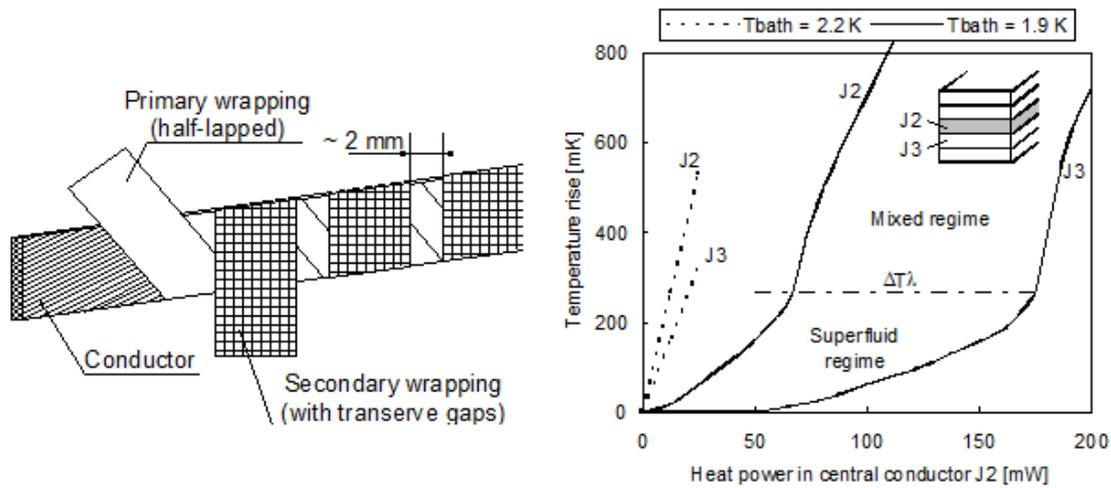

**Fig. 9:** The heat transfer across the polyimide-wrap insulation of a superconducting cable

The high thermal conduction in helium II can also be exploited to ensure quasi-isothermality of helium enclosures of limited spatial extension, such as the helium bath of a superconducting magnet under test. Knowledge of temperature changes at any point in the bath permits us to assess the enthalpy changes of the system, and thus to perform calorimetric measurements. This technique proves very convenient for measuring minute heat in-leaks [42] or substantial energy dissipation [43], such as produced by ramping losses or resistive transitions in superconducting magnets.

## 2.4 Forced-flow convection of pressurized superfluid helium

To overcome the limited conduction of helium II in long strings of cryogenic devices, the obvious issue is to create a forced circulation of the fluid in a cooling loop, thus relying on convective and advective heat transfer. One can then benefit from an additional control parameter, the net velocity imparted to the bulk fluid. In the following, we shall only discuss convection in channel diameters of technical interest; that is, typically greater than a few millimetres. The flow induced by a pressure gradient across a hydraulic impedance is then essentially determined by the bulk viscosity of the fluid. Assuming that internal convection between the components of the two-fluid model is independent of the net velocity reduces the problem to the behaviour of a flowing monophase liquid with high, non-linear thermal conductivity. The steady-state convective heat transport $\dot{Q}$ between two points 1 and 2 of the cooling loop is then given by the difference in enthalpy $H$ of the fluid flowing with a mass flow-rate $\dot{m}$:

$$\dot{Q} = \dot{m}(H_2 - H_1). \qquad (3)$$

An estimate of the potential advantage of forced convection over conduction can be made, using the same geometry and temperature boundary conditions as described in Section 2.2. Consider helium II pressurized at 100 kPa, flowing in a heated pipe of length 1 m and cross-section 1 cm$^2$, and assume that its temperature increases from 1.8 K at the pipe inlet to 1.9 K at the outlet. It is easy to show that for flow velocities above 0.2 m·s$^{-1}$, advective heat transport exceeds conduction.

The above calculation, however, neglects the pressure drop along the flow. A glance at the pressure–enthalpy diagram of helium (Fig. 10) reveals a positive Joule–Thomson effect [44]: the enthalpy of the fluid increases both with increasing temperature and pressure, so that an isenthalpic expansion results in a temperature increase. For example, pressurized helium II flowing across a pressure gradient of 50 kPa will warm up from 1.8 K to 1.9 K, in the absence of any applied heat load. The magnitude of this effect requires precise knowledge of the thermo-hydraulic behaviour of helium II, in order to validate its implementation in long cooling loops [45].

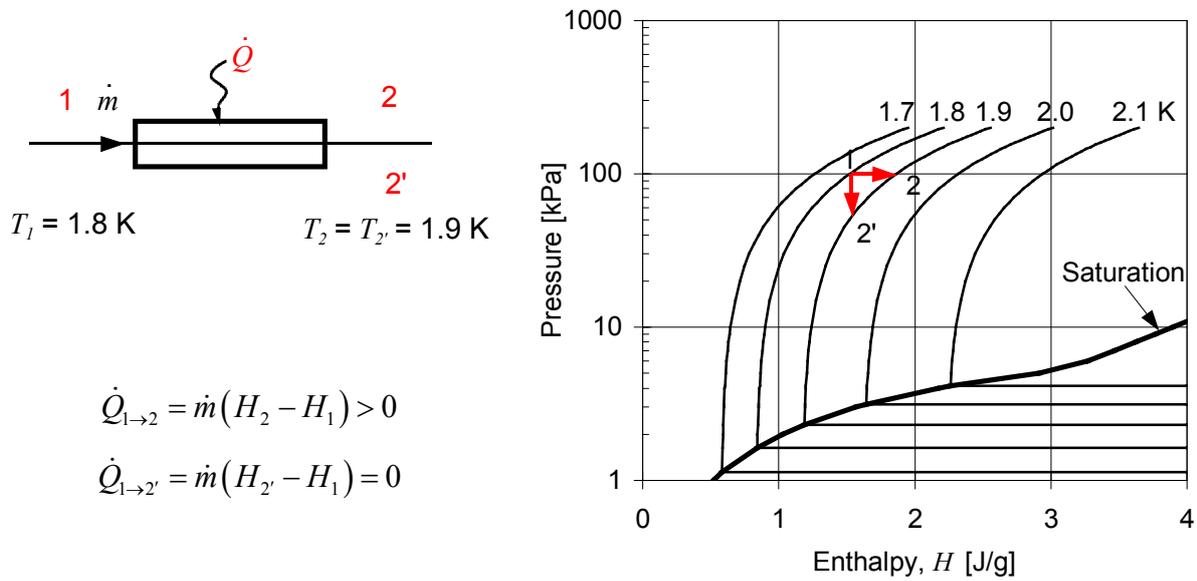

**Fig. 10:** The pressure–enthalpy diagram and forced-flow convection in superfluid helium

Following early work [46, 47], several experimental programmes have investigated the heated flow of pressurized helium II in pipes and piping components [48, 49], culminating with the 230-m long test loop at CEA Grenoble [50, 51], which gave access to high Reynolds numbers and extended geometries characteristic of accelerator string cooling loops. In parallel with that work, mathematical models were developed for calculating combined conductive and convective heat transport processes in complex circuits [52, 53], and were validated through experimental results. Pressure drop and heat transfer – both steady-state and transient – in flowing pressurized helium II may now be safely predicted for engineering purposes, using well-established laws and formulas.

The implementation of forced-flow cooling requires cryogenic pumps operating with pressurized helium II. Although most of the experimental work has been performed using positive displacement – that is, bellows or piston-pumps originally developed for helium I [54] – the thermo-mechanical effect, specific to the superfluid, may also be used for driving cooling loops by means of fountain-effect pumps [55–58]. In spite of their low thermodynamic efficiency [59], a drawback of limited relevance for using them as circulators, which usually have to perform low-pressure pumping work, fountain-effect pumps are light, self-priming and have no moving parts, assets of long-term reliability in, for example, on-board applications in space [60]. At higher heat loads, they have been considered [61] and tested [62] for forced-flow cooling of superconducting magnets: the overall efficiency of the process may then be improved by configuring the cooling loop so as to make use of the heat load of the magnet proper to drive the thermo-mechanical effect in the pump [63]. Practical implementation of these techniques for cooling superconducting magnets has, however, been limited, in particular by the need to provide other means of flow circulation during pre-cooling with normal helium above the lambda point.

### 2.5 Two-phase flow of saturated superfluid helium

The conductive and convective cooling systems described above both transport heat deposited or generated in the load over some distance through pressurized helium II, up to a lumped pressurized-to-saturated helium II heat exchanger acting as quasi-isothermal heat sink. This is achieved at the cost of a non-negligible – and thermodynamically costly – temperature difference, thus requiring the heat sink to operate several hundred millikelvins below the temperature of the load.

A more efficient alternative is to distribute the quasi-isothermal heat sink along the length of the accelerator string. In this fashion, the conduction distance – and hence the temperature drop – in the pressurized helium II is kept to a minimum, typically the transverse dimension of the device cryostat. This leads to the cooling scheme proposed for the LHC at CERN, schematized in Fig. 11: the superconducting magnets operate in static baths of pressurized helium II at around atmospheric pressure, in which the heat load is transported by conduction to the quasi-isothermal linear heat sink constituted by a copper heat exchanger tube, threading its way along the magnet string, and in which flowing two-phase saturated helium II gradually absorbs the heat as it vaporizes [11].

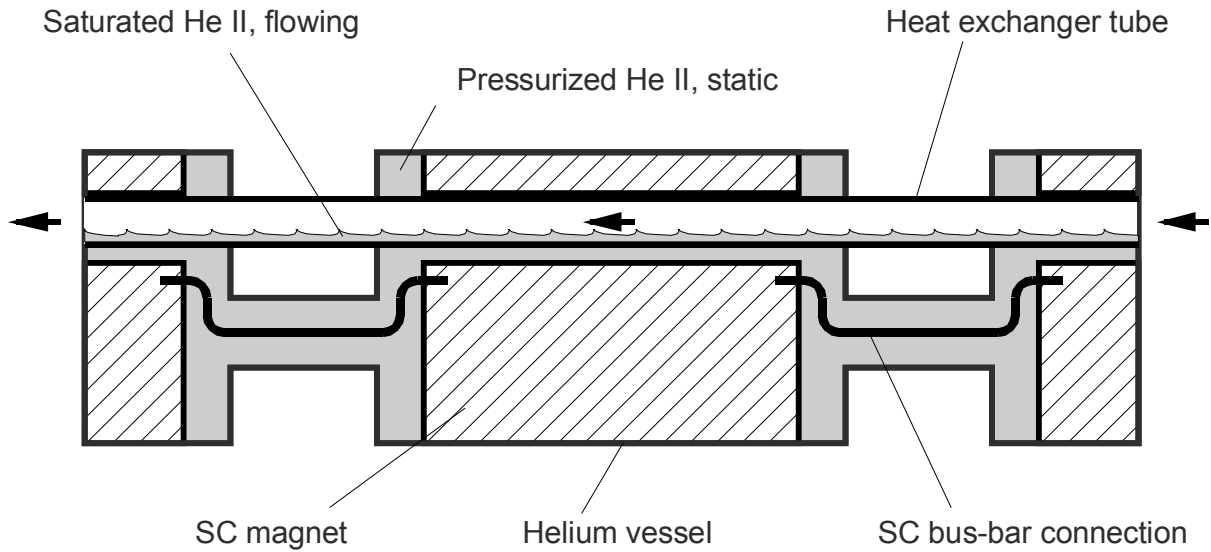

**Fig. 11:** The principle of the LHC superfluid helium cooling scheme

Although potentially attractive in view of its efficiency in maintaining long strings of magnets at quasi-uniform temperature, this cooling scheme departs from the well-established wisdom of avoiding long-distance flow of two-phase fluids at saturation, particularly in horizontal or slightly inclined channels. Moreover, no experimental data was originally available on flowing saturated helium II, and very little for other cryogenic fluids in this configuration. Following the first exploratory tests [64], which demonstrated the validity of the concept on a reduced geometry, a full-scale thermo-hydraulic loop [65] permitted us to establish the stability of horizontal and downward-sloping helium II flows, to observe partial (but sufficient) wetting of the inner surface of the heat exchanger tube by the liquid phase, due to flow stratification, and to address process-control issues and develop strategies for controlling uniformity of temperature at strongly varying applied heat loads, given the low velocity of the liquid phase. As long as complete dry-out does not occur, an overall thermal conductance of about 100 $W·m^{-1}·K^{-1}$ can be reproducibly observed across a DN40 heat exchanger tube, made of industrial-grade deoxidized phosphorus copper.

Once the wetting of the inner surface of the tube is guaranteed, the heat transfer from the pressurized to the saturated helium II is controlled by three thermal impedances in series: solid conduction across the tube wall, and Kapitza resistance at the inner and outer interfaces between tube wall and liquid (Fig. 12). While the former can be adjusted, within technological limits, by choosing the tube material and wall thickness, the latter, which finds its origin in the refraction of phonons at the liquid/solid interfaces and is thus strongly temperature dependent, usually dominates below 2 K [66].

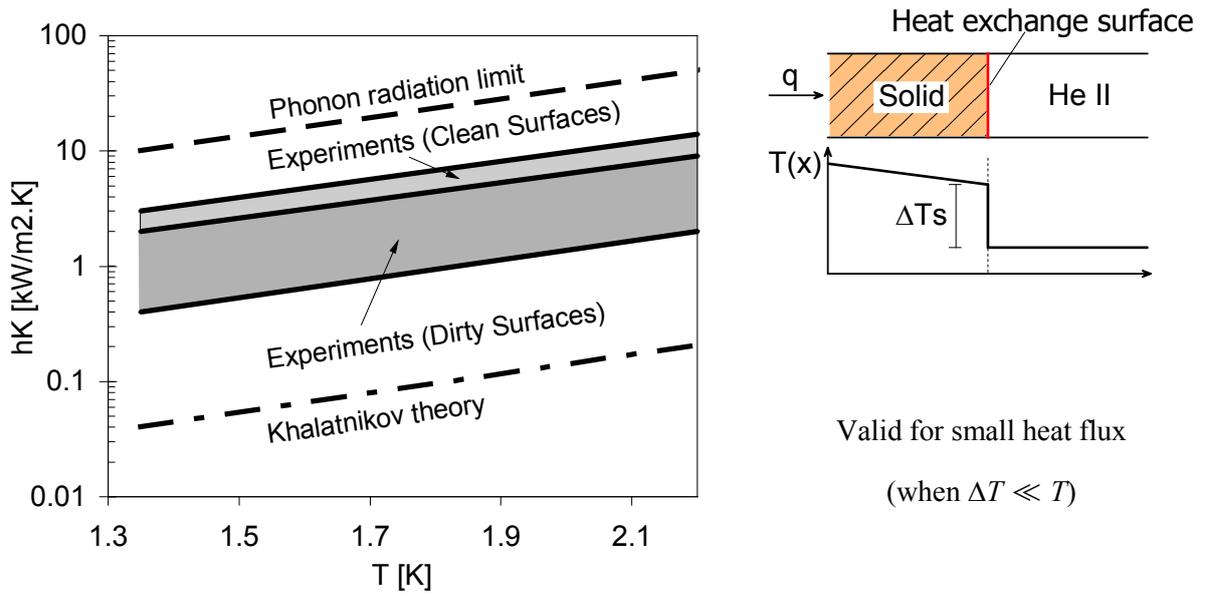

**Fig. 12:** The Kapitza conductance at a copper/helium II interface

The final validation of the two-phase helium II flow cooling scheme for the LHC has been performed successfully on a 100-m long test string, equipped with full-scale prototype cryomagnets, operated and powered in nominal conditions [67, 68]. At varying heat loads exceeding 1 W·m$^{-1}$, all magnets in the string were maintained in a narrow range of temperature, a few tens of millikelvins above the saturation temperature of the flowing helium II. Thermal buffering provided by the pressurized helium II baths contributed to limiting temperature excursions, at the cost of introducing strong non-linearities and time delays in the system, which must be coped with by elaborate, robust process control [69, 70]. As a complement to that applied work focused towards LHC, more fundamental experimental studies have been conducted on specially instrumented test loops at CEN-Grenoble, comprehensively equipped with diagnostics and a transparent section for visual observation and interpretation of the flow patterns [71–73]. As long as the vapour velocity remains sufficiently low to maintain stratified flow (up to a few metres per second), the engineering design of such a cooling scheme rests on a few simple sizing rules [74]. At higher vapour velocity, entrainment and atomization effects complicate the flow pattern and impact on the heat transfer and pressure drop [75].

This type of cooling scheme may also be used for extracting much higher linear heat loads, typically about 10 W·m$^{-1}$, as present in the low-beta quadrupoles of the high-luminosity insertions of the LHC [76, 77], at the expense of a larger-diameter heat exchanger tube to limit the saturated vapour velocity and thus preserve flow stratification.

## 3    Refrigeration cycles and equipment

The properties of helium at saturation (see Fig. 3) make it necessary to maintain an absolute pressure below 1.6 kPa on the heat sink of a 1.8 K cryogenic system. Bringing the saturated vapour up to atmospheric pressure thus requires compression with a pressure ratio exceeding 80; that is, four times that of refrigeration cycles for 'normal' helium at 4.5 K. Figure 13 shows the basic scheme for refrigeration below 2 K. A conventional refrigerator produces liquid helium at 4.5 K, later expanded down to 1.6 kPa in a Joule–Thomson expansion stage. The gaseous helium resulting from liquid vaporization is compressed above atmospheric pressure and eventually recovered by the 4.5 K refrigerator. We will therefore start by presenting the Joule–Thomson expansion stage.

Three types of cycles, sketched in Fig. 13, can be considered [78, 79] for producing refrigeration below 2 K:

- the 'warm' compression cycle, based on ambient-temperature sub-atmospheric compressors;
- the 'cold' compression cycle, based on multistage cold compressors all the way up to atmospheric pressure; and
- the 'mixed' compression cycle, based on a combination of cold compressors in series with ambient-temperature sub-atmospheric compressors.

We will then proceed to discuss the thermodynamics and machinery for these three types of cycle.

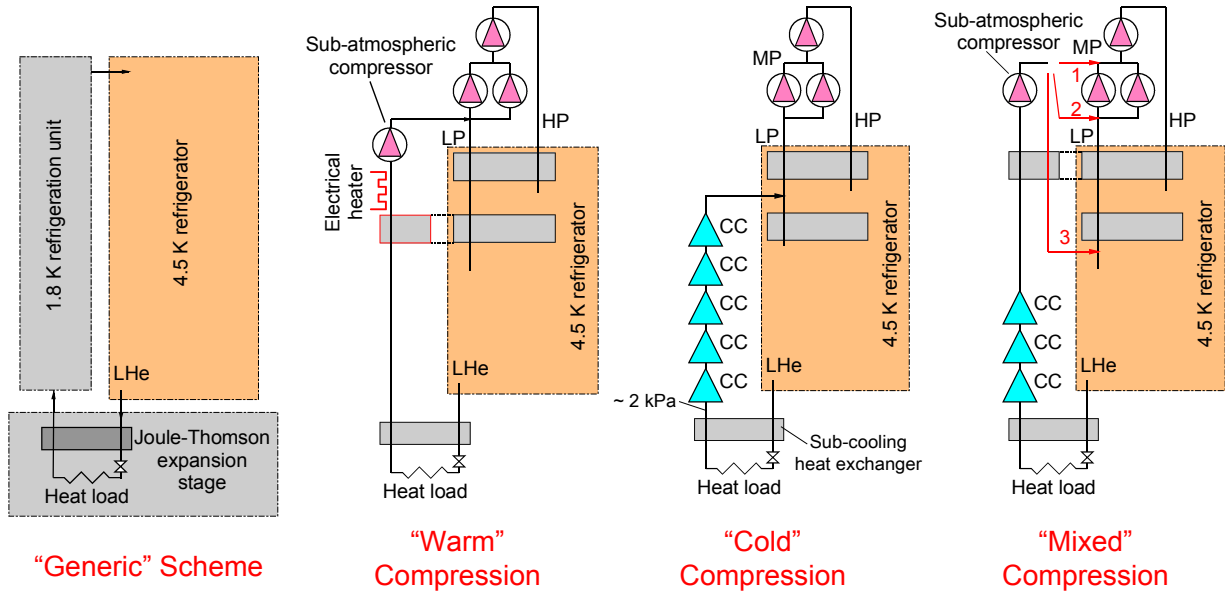

**Fig. 13:** Generic process cycles for refrigeration below 2 K

## 3.1 The Joule–Thomson expansion stage

The efficiency of the Joule–Thomson expansion of liquid helium, say from 0.13 MPa and 4.5 K, down to 1.6 kPa and 1.8 K, can be notably improved if the liquid is previously sub-cooled by the exiting very-low-pressure vapour (Fig. 14). This is performed in a counter-flow heat exchanger, sub-cooling the incoming liquid down to 2.2 K by enthalpy exchange with the very-low-pressure saturated vapour. This heat exchanger has to produce a limited pressure drop, particularly in the very-low-pressure stream. A maximum pressure drop of 100 Pa is generally acceptable, corresponding to a few per cent of the absolute saturation pressure. The design of such heat exchangers for large flow-rate [80] is not straightforward, and their qualification impractical. As a consequence, the LHC cryogenic system features several hundred small-sized (5–20 g·s$^{-1}$) heat exchangers, distributed around the ring. This avoids the transportation of sub-cooled helium over long distances, saving one header in the ring distribution line. It also permits qualification and reception testing of the heat exchangers on a test stand of reasonable size. Following prototyping, technical validation of different solutions [81, 82] and commercial selection, these heat exchangers have been series produced by industry and are now in operation in the LHC tunnel.

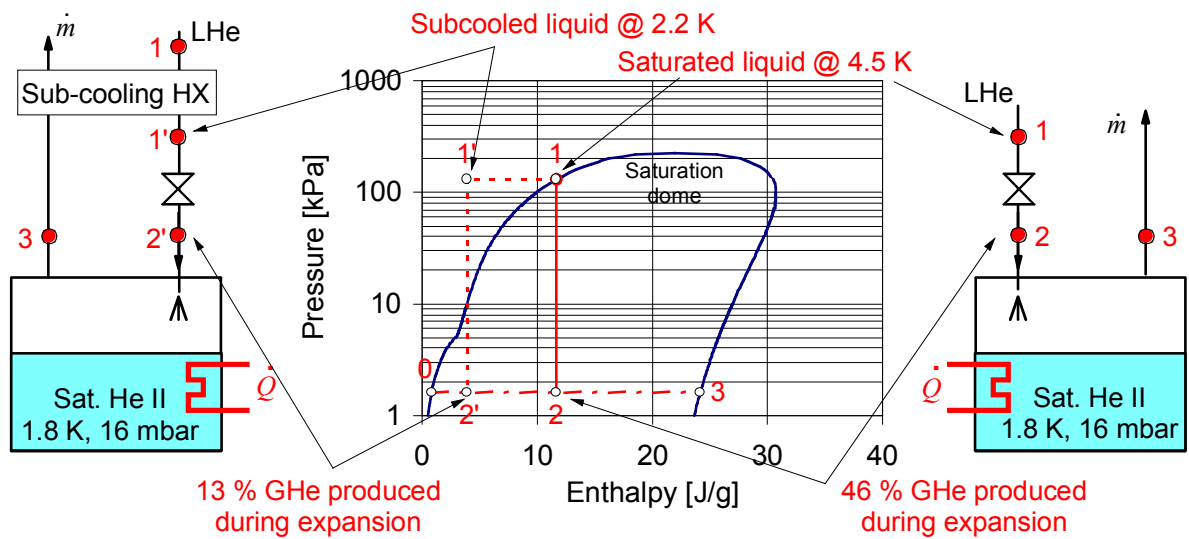

**Fig. 14:** The efficiency of the Joule–Thomson expansion

### 3.2 The 'warm' compression cycle

For low-power refrigeration – for example, in small laboratory cryostats – this is achieved by means of standard Roots or rotary-vane vacuum pumps (Fig. 15), handling the very-low-pressure gaseous helium escaping from the bath after it has been warmed up to ambient temperature through a heat exchanger and/or an electrical heater. This technology may be pushed to higher flow-rates using liquid-ring pumps, adapted for processing helium by improving the tightness of their casing and operating them with the same oil as that of the main compressors of the 4.5 K cycle [83], or oil-lubricated screw compressors operating at low suction pressure (Fig. 15). In any case, compression at ambient temperature is hampered by the low density of the gaseous helium, which results in large volume flow-rates and thus requires large machinery, as well as in costly, inefficient heat exchangers for recovering the enthalpy of the very-low-pressure stream.

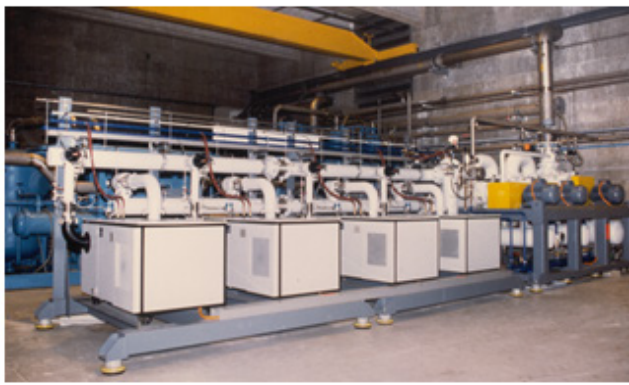
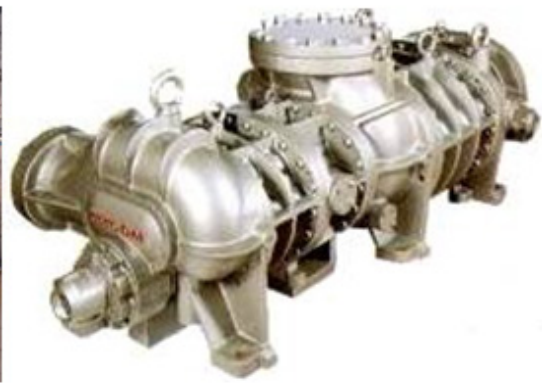

(a)      (b)

**Fig. 15:** Sub-atmospheric compressors: (a) a combination of Roots and rotary-vane vacuum pumps; (b) the compound screw.

All these compressors are positive-displacement machines having volumetric characteristics. Screw compressors are routinely used in helium refrigeration and their implementation in a 1.8 K cycle therefore follows from current practice. Special attention, however, has to be paid to the protection against air in-leaks: in particular, the motor shaft and its rotary sealing must be located on the discharge side to operate above atmospheric pressure.

A first limit to the use of sub-atmospheric screw compressors stems from volumetric flow requirements: the biggest available machines have a swept volume of about 4600 m$^3 \cdot$h$^{-1}$, so that higher flow-rates require parallel arrays. Moreover, the isothermal efficiency – defined as the ratio of isothermal compression work to the effective compression work of the machine – decreases markedly with the suction pressure, as shown in Fig. 16, thus precluding their use at very low pressure in efficient process cycles.

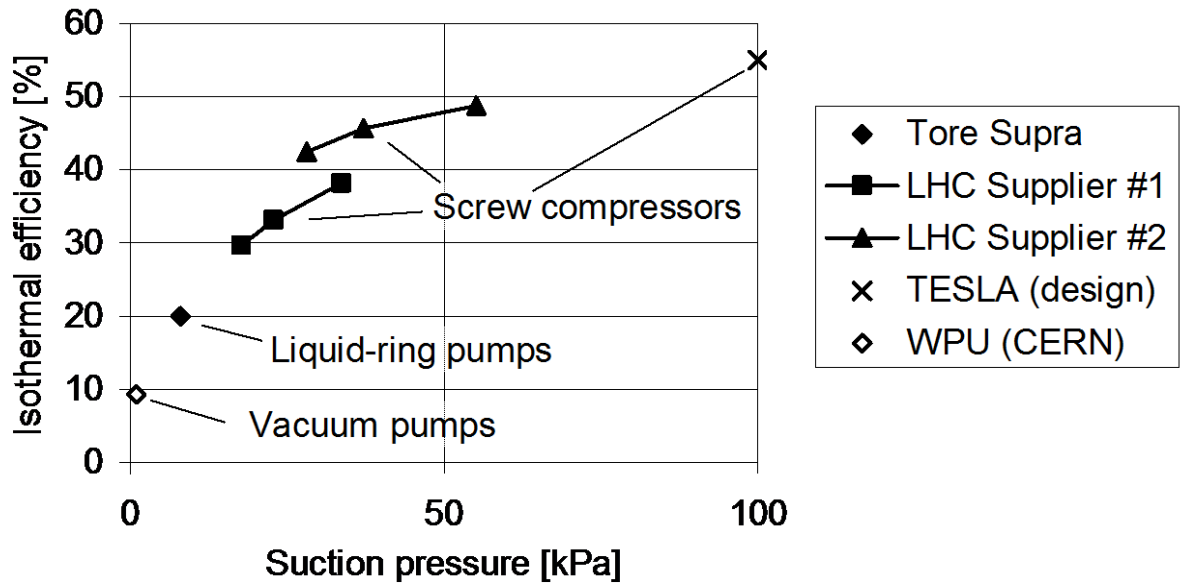

**Fig. 16:** The isothermal efficiency of warm sub-atmospheric compressors

### 3.3 The 'cold' compression cycle

The alternative process is to perform compression of the vapour at low temperature; that is, at its highest density. The pumps and recovery heat exchangers get smaller in size and less expensive, but the compression work is then injected in the cycle at low temperature, so that the inevitable irreversibilities have a higher thermodynamic weight. Moreover, the pumping machinery that handles the cold helium must be non-lubricated and non-contaminating, which seriously limits the choice of technology. Hydrodynamic compressors, of the centrifugal or axial–centrifugal type, have been used in large-capacity systems [84]. Their pressure ratio, limited to 2–3.5 per stage, however, makes it necessary to arrange them in multistage configurations [85, 86], thus narrowing the operational range of the system, in particular for start-up or off-design modes.

Depending on the operating temperature (2.0 K or 1.8 K), the 'cold' compression cycle requires at least four or five stages in series in order to perform the overall pressure ratio of 45–80. The compressed helium is directly returned to the cold low-pressure (LP) stream of the 4.5 K refrigerator.

The main drawback of this cycle concerns turndown capability. The cold compressor set has to guarantee the same pressure ratio for any load. A typical operating field for hydrodynamic compressors (Fig. 17) displays the pressure ratio as a function of the reduced flow $m^*$ and the reduced speed $N^*$. The working area is limited on the left-hand side by the stall line, on the right-hand side by the choke line, and on top by the maximum rotational speed of the drive. At constant pressure ratio, the compressor can handle a flow reduction of only about 20% before reaching the stall line. Below 80% of nominal, additional vapour generation by electrical heating must be used to compensate for the load reduction. Such a cycle is therefore not very compliant to turndown, and its operating cost is not optimized for part-load operation.

This led CERN to conduct, in view of the LHC project, a R&D programme on cold compressors, procuring from specialized industry three prototype hydrodynamic compressors of different designs [87–90] to investigate critical issues such as drive and bearing technology, impeller and diffuser hydrodynamics, and mechanical and thermal design, as well as their impact on overall efficiency [91]. The choices eventually retained for the LHC series machines [92–94] are three-phase electrical induction motor drives working at room temperature, with a rotational speed varying from 200 to 700 Hz, active magnetic bearings working at room temperature, axial–centrifugal (three-dimensional) impellers, and fixed-vane diffusers (Figs. 18 and 19).

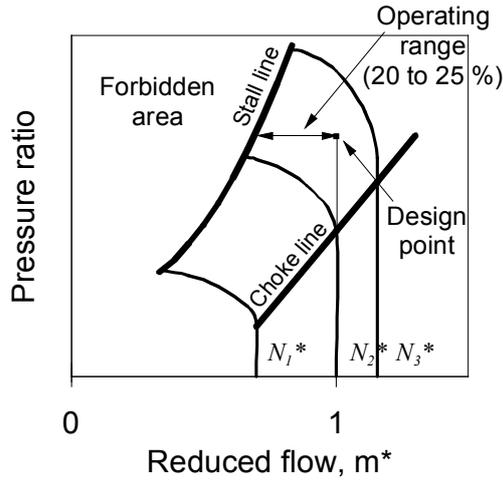

$$m^* = \frac{\dot{m}}{\dot{m}_0} \times \sqrt{\frac{T_{in}}{T_{in_0}}} \times \frac{P_{in_0}}{P_{in}}$$

and

$$N^* = \frac{N}{N_0} \times \sqrt{\frac{T_{in_0}}{T_{in}}}$$

with: $\dot{m}$ = mass-flow
$T_{in}$ = inlet temperature
$P_{in}$ = inlet pressure
$N$ = rotational speed
subscript 0 = design condition

**Fig. 17:** The typical operating field of a hydrodynamic compressor

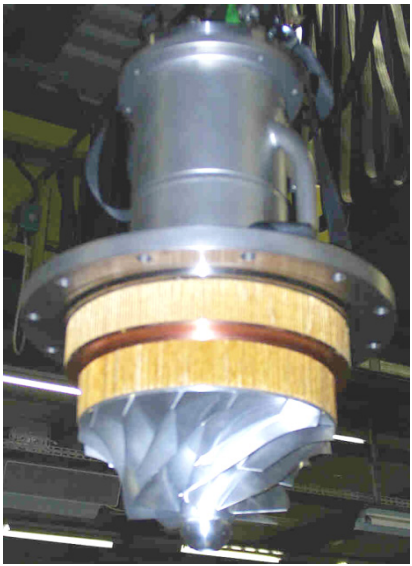
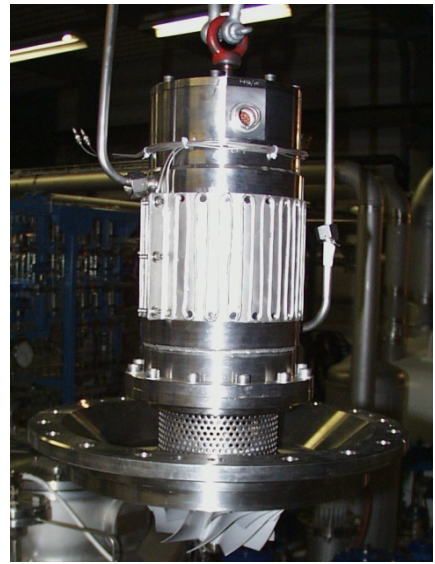

**Fig. 18:** Axial–centrifugal cold compressor cartridges for the LHC

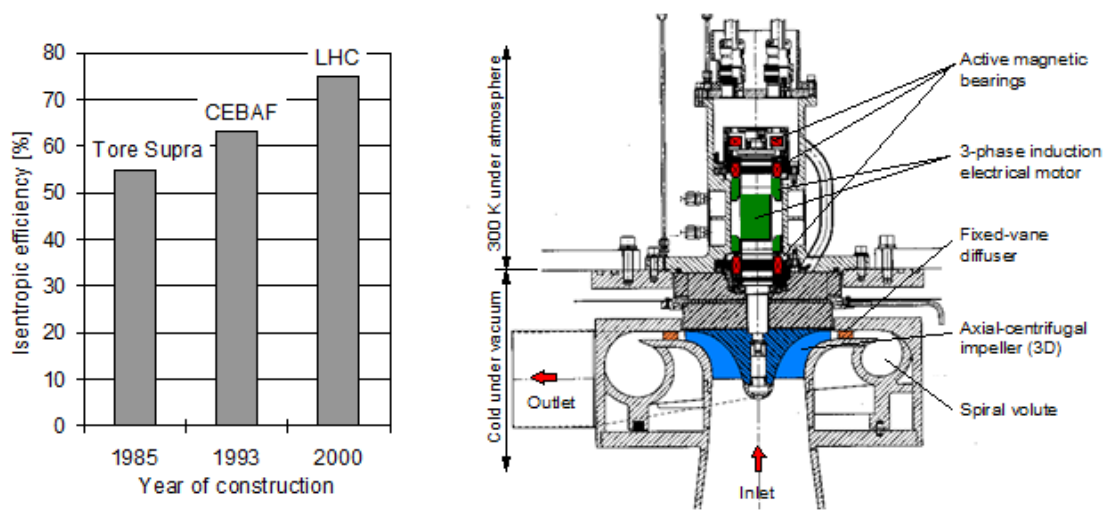

**Fig. 19:** The isentropic efficiency and a typical cross-section of a cold compressor

The thermodynamic efficiency of cold compressors, determined by hydrodynamic design as well as by the limitation of internal leakage and of heat in-leaks along the drive shaft, has significantly improved (Fig. 19). Here, the relevant estimator is the isentropic efficiency, defined as the ratio of the compression work in the adiabatic, reversible case, to the real one. Recent machines reach 75% isentropic efficiency at their design point.

## 3.4   'Mixed' compression cycles

For large systems, oil liquid-ring pumps or lubricated screw compressors may be used in series with cold compressors, in 'mixed' compression cycles. Cold compressors are well suited for the lower stages, while the presence of volumetric machines in the upper stages permits independent adjustment of the flow-rate or wheel inlet conditions, thus improving load adaptation [95, 96].

In 'mixed' compression cycles, the number of cold compressor stages can be reduced to three, depending on the swept volume and the number of warm sub-atmospheric machines. The compressed helium can be returned to the 4.5 K refrigerator at different levels.

- At the warm medium-pressure (MP) side (connection #1 in Fig. 13(d)). This requires the use of screw compressors having a sufficient built-in pressure ratio. In this case, the enthalpy of the gas at the outlet of the cold compressors has to be recovered by the heat exchangers of the 4.5 K refrigerator. The main advantage of this solution is that the same oil-removal and final cleaning systems can be used for the warm sub-atmospheric compressors and for the booster stages of the 4.5 K refrigerators, thus minimizing the investment cost of the system.

- At the warm low-pressure (LP) side (connection #2 in Fig. 13(d)). This solution is compatible with the use of either screw compressors or liquid-ring pumps. The enthalpy of the cold gas at the outlet of the cold compressors also has to be recovered by the heat exchangers of the 4.5 K refrigerator. In this case, the warm sub-atmospheric stage requires its own oil-removal system.

- At the cold low-pressure side (connection #3 in Fig. 13(d)). This is required when the enthalpy of the cold gas at the outlet of the cold compressors cannot be recovered by the heat exchangers of the 4.5 K refrigerators (the LHC case) [92]. In this case, the warm sub-atmospheric stage requires its own oil-removal and final cleaning system (coalescing filters and charcoal adsorbers), increasing the investment cost.

The main advantage of the 'mixed' cycle resides in its turndown capability. With sub-atmospheric compressors having volumetric characteristics, the pressure at the outlet of the cold compressors decreases linearly with the flow-rate; that is, if the temperature and rotational speed do not change, the reduced flow-rate $m^*$ stays constant, thus keeping the working point fixed in the operating field. Such a cycle can then handle a large dynamic range – for example, a value of 3 for the LHC – without any additional electrical heating. Moreover, the total pressure ratio of the cold compressor train is lowered and the speed of some machines can then be reduced, thus decreasing the total compression power and operating cost, and improving the overall efficiency of the cycle.

Another operational advantage concerns the possibility of maintaining the load in cold standby with the cold compressors freewheeling and all compression performed, though at much reduced flow, by the warm machines. This mode allows repair or exchange of a cold-compressor cartridge without emptying the helium from the system. In addition, the load adaptation provided by the warm volumetric machines proves very useful during transient modes such as cool-down and pump-down, in which the cold compressors operate far from their design conditions.

The only drawback of this cycle concerns the risk of air in-leaks due to the presence of sub-atmospheric circuits in air. Helium guards are recommended to prevent pollution of the process helium [97].

## 3.5 The application range of low-pressure helium compression techniques

The practical ranges of application of the different techniques are shown in Fig. 20, setting a *de facto* limit for warm compression above 20 000 m$^3$·h$^{-1}$, or typically 300 W at 1.8 K. The diagram also illustrates the large span of refrigeration power and the diversity of projects using superfluid helium. The investment and operating costs of large superfluid helium refrigeration systems can be assessed from basic thermodynamics and practical scaling laws derived from recent experience [98], thus providing input for the technical–economical optimization of such systems.

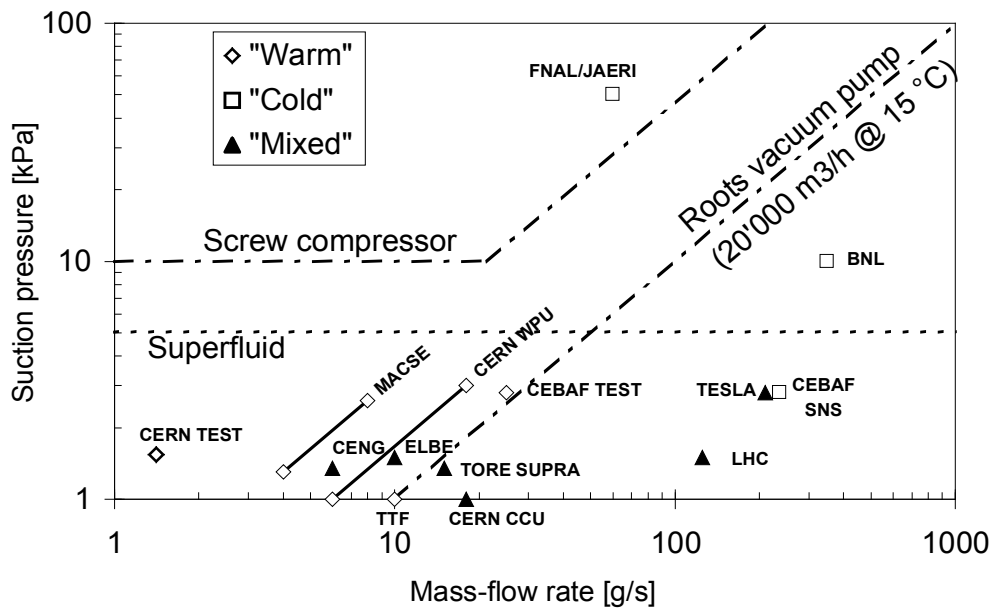

**Fig. 20:** The range of application of low-pressure helium compression techniques

# 4 Other technological aspects

The preceding sections have addressed the main specific aspects of superfluid helium technology, namely the properties of helium II, cooling methods, and refrigeration below 2 K. There is, of course, a lot more to the design and construction of a fully operational helium II cryogenic system for cooling superconducting devices in a particle accelerator, but it is less specific – if at all – to superfluid helium, and can be dealt with by adopting the design rules and good practice required by more conventional liquid helium technology. Still, three aspects are singled out in what follows, as they are often the subject of questions asked by newcomers to the field.

## 4.1 Helium II to vacuum leak-tightness

The low viscosity of helium II and the thermo-mechanical effect it displays across 'superleaks' often raise the question of special techniques for designing and building tight vessels, pipelines and enclosures for helium II systems. Experience has shown that the state-of-the-art design and construction rules applicable for cryogenic equipment operating with normal liquid helium are sufficient to ensure helium II leak-tightness. This may be explained by the fact that a minute crack in the wall of the helium II vessel of a cryostat, surrounded by insulation vacuum, is physically different from a 'superleak'; that is, porosity across a plug separating two helium II enclosures. In the former case, the leaking helium will vaporize when it reaches saturation pressure some way along the crack and the leakage rate will eventually be controlled by the flow of vapour downstream – as it would be for a normal helium leak.

The practical approach used for ensuring leak-tightness of the LHC helium II enclosure – essentially the 'cold masses' of the superconducting magnets and their interconnections – was to enforce an all-welded austenitic stainless steel construction, using automatic welding to reduce human error. While TIG welding was used for low-thickness material – for example, the cryomagnet interconnections [99] – a combination of STT (root pass) and MAG (filling passes) was used for the 10-mm thick AISI 316 LN wall of the magnet 'cold mass' [100]. Local and global leak detection following the pressure test was systematically performed as an integral part of the quality assurance procedure [101].

Specific strategies were applied in exceptional cases when an all-welded construction could not be used. The copper heat exchanger tubes running along the LHC magnet strings [102] have brazed end-sleeves of austenitic stainless steel, so that they can be TIG-welded to each other when interconnecting the cryomagnets; by design, the brazed joint is located, upon assembly of the magnet, inside the all-welded helium enclosure, so that it only has to provide helium-to-helium tightness. Another case was cold reception tests, for which the cryomagnets were connected to the test benches by demountable, double-ring metal seals, providing a vacuum-monitored and spectrometer leak-detected guard volume.

## 4.2 Air to sub-atmospheric helium leak-tightness

Even when operating with pressurized superfluid helium, there will be sections of the helium circuits below atmospheric pressure: the cold parts of these circuits, vacuum insulated for thermal reasons, therefore also benefit from a vacuum guard that prevents any contamination of the helium circuit by air in-leak from the atmosphere. The room-temperature parts of the circuits are usually in contact with the atmosphere, and any leak will result in contamination. In welded circuits, such leaks may occur across demountable seals, instrumentation (e.g. pressure sensors), or non-tightly closing relief valves. Corrective measures against such leaks, as schematized in Fig. 21, make use of helium guard volumes: the helium guard may be a dedicated vessel enclosing the potentially leaky component, or the space between double, concentric seals.

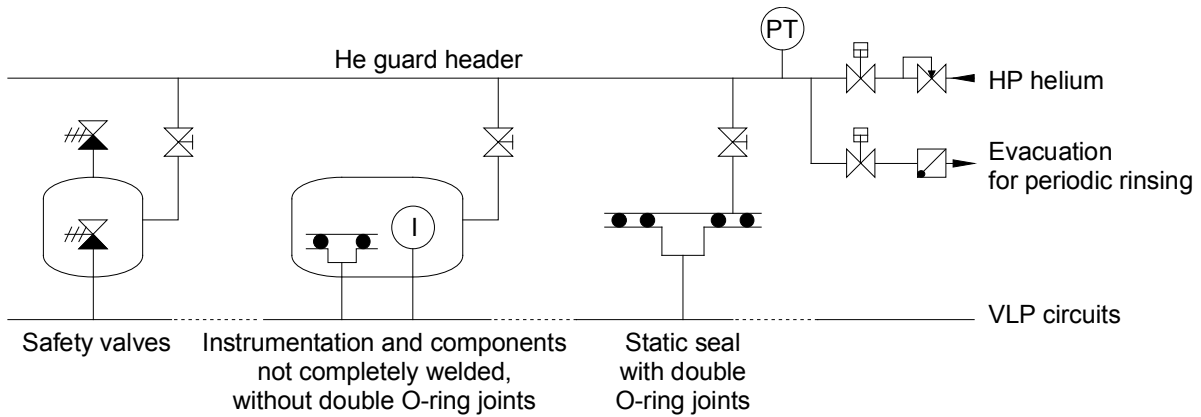

**Fig. 21:** Helium guards against air in-leaks into sub-atmospheric helium circuits

Concerning the protection of cryogenic control valves operating at sub-atmospheric pressure, such as the Joule–Thomson expansion valves mentioned in Section 3.1, it is advisable to install these valves with the higher-pressure side not under the seat as recommended by general practice but, rather, on the stem side where the sealing interface with the atmosphere occurs; an imperfect sealing would then result in helium leakage to the atmosphere, rather than air in-leak into the helium circuit. For cryogenic valves that have sub-atmospheric pressure on both sides, double sealing with an intermediate helium guard is required on the stem interface with the atmosphere, as an application of the general protection principles described above.

## 4.3 Precision thermometry

Low-temperature precision thermometry is always challenging, whether it uses the primary thermometers of the International Temperature Scale ITS-90 [103] or more practical secondary thermometers that have to be calibrated against them. In a cryogenic process at superfluid helium temperature, errors in temperature measurement may result in significant irreversibilities; for example, by requiring an increase of the temperature gradients for transporting heat so that they include error bars. As an example, the total temperature gradient along a 3.3 km long sector of the LHC, from the warmest magnet at the end of the sector to the refrigeration plant at 1.8 K, is only 100 mK, including all heat transfer processes and uncertainties; allowing a 10% uncertainty on this value requires the thermometers to have an absolute precision better than ±5 mK in this temperature range.

The primary thermometer of the ITS-90 in the superfluid helium domain is the helium vapour pressure curve: measurement of temperature on a saturated bath is therefore reduced to that of pressure, provided that all sources of parasitic heads in the measurement system – for example, hydrostatic – are correctly identified and compensated. In other cases, one generally uses resistive temperature sensors, usually based on semiconductor materials showing high non-linearity. Once the selected sensors have demonstrated their stability over time, thermal cycles, and variation of environmental conditions – for example, ionizing radiation – they need to be individually calibrated to achieve precision [104]. The calibration data must then be reduced to be practically implemented in readout electronics and/or monitoring software.

Finally, it must be remembered that the best thermometer can only measure its *own* temperature! This is particularly important for thermometers operating in vacuum, for which good thermal coupling to the object to be measured, protection against parasitic heat in-leaks, whether conductive (along readout wires) or radiative, and limitation of internal Joule heating by the readout current are essential to the quality of the measurement. Thermometric blocks integrating all these functions [105] have been successfully developed, to be installed in the field by non-specialized personnel.

## 5  Conclusion

The operation of superconducting devices in particle accelerators below 2 K, using superfluid helium as a technical coolant, has now become state-of-the-art, as exemplified by the excellent operational records of CEBAF at the Thomas Jefferson National Accelerator Facility, SNS at the Oak Ridge National Laboratory, and the LHC at CERN. The specific aspects of superfluid helium technology – addressed in this chapter – can be combined with standard cryogenic practice to design, build, and operate complete helium II systems of industrial size. The superfluid-helium cooled particle accelerator projects in construction or under study, such as the European Spallation Source (ESS) and the International Linear Collider (ILC), however, represent major challenges and opportunities for further progress, in view of their large size, their complexity, and the quest for reliability and efficiency.